\documentclass[12pt,preprint]{aastex}

\usepackage{epsfig}
\usepackage{bm}

\shorttitle{Wavelets and WMAP non-Gaussianity}
\shortauthors{Mukherjee \& Wang}

\begin{document}

\title{Wavelets and WMAP non-Gaussianity}
\author{Pia~Mukherjee \& Yun~Wang}
\affil{Department of Physics \& Astronomy, Univ. of Oklahoma,
                 440 W Brooks St., Norman, OK 73019;
                 email: pia,wang@nhn.ou.edu}

\begin{abstract}
We study the statistical properties of the 1st year WMAP data on different
 scales using the spherical mexican hat wavelet transform. 
Consistent with the results of Vielva et al. (2003) we find a deviation
 from Gaussianity in the form of kurtosis of wavelet coefficients on 
$3-4^\circ$ scales in the southern Galactic hemisphere. This paper 
extends the work of Vielva et al. as follows. We find that the non-Gaussian
 signal 
shows up more strongly in the form of a larger than expected number of cold
 pixels and also in the form of scale-scale correlations amongst wavelet 
coefficients. We establish the robustness of the non-Gaussian signal
 under more wide-ranging assumptions regarding the Galactic mask applied
 to the data and the noise statistics. This signal is unlikely to be due
 to the usual quadratic term parametrized by the non-linearity parameter
 $f_{NL}$. We use the skewness of the spherical mexican hat
 wavelet coefficients to constrain $f_{NL}$ with the 1st year WMAP data.
 Our results constrain $f_{NL}$ to be $50\pm 80$ at 68\% confidence,
 and less than 280 at 99\% confidence.
\end{abstract}

\keywords{Cosmology: Cosmic Microwave Background, methods: statistical}

\section{Introduction}

The current cosmological model assumes Gaussian initial conditions, created by inflation. 
This assumption regarding the nature of primordial density perturbations can be verified 
by studying the distribution of temperature fluctuations in the cosmic microwave background 
(CMB). While the simplest inflationary models predict Gaussian primordial perturbations, 
there are other models of inflation, such as those involving multiple scalar fields, 
features in the inflaton potential or phase transitions, that could give rise to detectable 
non-Gaussianity. Hence studies of Gaussianity help distinguish between different early 
universe scenarios. Gaussianity is also a key underlying assumption of CMB data analysis wherein the angular power spectrum fully specifies its statistical properties, and must be tested. Non-Gaussianity can also be associated with secondary anisotropies 
in the CMB, or with foreground contamination and systematic effects.

Prior to the release of WMAP data there was no clear evidence of cosmological 
non-Gaussianity. Since the release of 1st year WMAP data, a number of tests of 
non-Gaussianity have been performed, with somewhat differing results. Each statistic 
is sensitive to a different kind of non-Gaussianity, hence there is need for a wide 
variety of tests. Komatsu et al. (2003) use an optimized test based on the bispectrum, 
as well as Minkowski functionals, while Colley \& Gott (2003) study the genus, and 
both groups report consistency with Gaussianity. Gaztanaga \& Wagg (2003) do a 
3-pt angular correlation function analysis and find consistency with Gaussianity as well. 
Chiang et al. (2003) perform a study of the phases of spherical harmonics and find 
some evidence for non-Gaussianity at high multipoles. Copi et al. (2003) find some 
evidence for low $l$ correlations and deviation from isotropy. Park (2003) find a 
large difference between the genus amplitudes of the northern and southern hemispheres 
and a positive genus asymmetry in the southern hemisphere. Eriksen et al. (2003a) 
compute the 2 and 3-pt correlations and report a significant north-south asymmetry; 
Eriksen et al. (2003b) use Minkowski functionals and find a significant genus in 
the northern hemisphere and again indications of north-south asymmetry.  
Hansen et al. (2004) use local curvature and find non-Gaussianity/asymmetry in the data on scales of a few degrees. Gurzadyan et al. (2004) find ellipticity in the temperature anisotropy features in the data, consistent with what was found previously in BOOMERang data.
Vielva et al. (2003; hereafter V03) report a non-Gaussian signal in the southern hemisphere at high significance in the form of kurtosis on $\sim 4^\circ$ scales using the 
spherical mexican hat wavelet transform on WMAP data. Some of the detections of non-Gaussianity and/or asymmetry
thus far reported in the WMAP data are at the level of 99\% or greater.

Wavelet transforms are useful tools in non-Gaussianity studies because they 
enable the signal on the sky to be studied on different scales, with simultaneous 
position localization, so that the obscuring effects of the central limit theorem, 
that can exist in both real and Fourier spaces, are reduced. With wavelets any 
non-Gaussian detection can be localized on the sky in scale and position, so that 
its nature and source can be better determined. Planar wavelets have been used 
in Gaussianity studies of the CMB by Pando et al. (1998), Hobson et al. (1999) 
and Mukherjee et al. (2000), while Barreiro et al. (2000) use the spherical Haar wavelet, 
and Cay\'on et al. (2001,2003), Mart\'inez-Gonz\'alez et al. (2002), and V03
use the spherical mexican hat wavelet (SMHW). Wavelet methods have been compared with 
other pixel or Fourier based methods in Hobson et al. (1999), Aghanim et al. (2003), 
Cabella et al. (2004), and the performance of isotropic as well as highly anisotropic 
multi-scale bases in distinguishing between different sources of non-Gaussianity 
in the CMB has been studied in Starck et al. (2003).

In this paper we use the spherical mexican hat wavelet transform to probe non-Gaussianity in the WMAP data. We extend the work of V03 by performing new multiple tests of the robustness of the non-Gaussian signal. We also look at the non-Gaussianity in terms of an excess in the number of cold pixels, and in terms of scale-scale correlations amongst wavelet coefficients. Further, we place constraints on a popular form of non-Gaussianity (a quadratic term in the curvature perturbations parametrized by the non-linearity parameter $f_{NL}$). This paper is organized as follows. 
In \S 2, we present results from using the SMHW transform on WMAP data. 
Whilst confirming the results of V03, we perform new
multiple tests of the robustness of the non-Gaussianity signal in the kurtosis spectrum (a) through the use of different (extended) masks, and (b) relaxing the assumption of a simplified noise model. We find that the signal shows up even more significantly in the form of the number of cold pixels (or coefficients).
In \S 3, we examine scale-scale correlations amongst the wavelet coefficients. We find significant deviations from Gaussianity, a corroboration of the signal detected and described in \S 2.
In \S 4 we obtain constraints on the non-linearity parameter $f_{NL}$.
We conclude in \S 5.

\section{Skewness and Kurtosis of Wavelet Coefficients}

A non-Gaussianity detection in the 1st year WMAP data was reported by V03. 
Applying the SMHW transform to the Q-V-W coadded data, and computing the skewness and 
kurtosis of the wavelet coefficients over scales ranging from about 10 arcmins to 10 
degrees, they found that the kurtosis of wavelet coefficients on scales $\sim4^\circ$ was 
too high at the 99.9\% confidence level. It was found that the excess kurtosis was in the 
southern hemisphere, while the kurtosis signal in the northern hemisphere was consistent 
with Gaussianity. The signal was shown to be independent of frequency.

It is important to determine whether the significance of this non-Gaussianity detection is affected by systematic effects, such as the choice of mask, or simplified assumptions
about noise. Since the spherical mexican hat wavelet transform is
a sensitive probe of non-Gaussianity, we use it
to perform an independent analysis of the 1st year WMAP data.
The basic steps followed in the analysis are as follows: 
starting with the foreground cleaned Q-V-W coadded data, 
we bring the map down in resolution to Healpix $n_{side}=256$, 
apply the $kp0$\footnote{By $kp0$ we mean the $kp0$ mask without sources.} 
mask, perform SMHW transforms\footnote{See Appendix for further details.} 
to obtain wavelet coefficients corresponding to the different scales $R$ 
(also setting the monopole and dipole of the map to zero here), apply 
appropriately extended versions of the mask to the wavelet coefficients of 
each scale to exclude coefficients contaminated by the mask and known point sources, and compute the 
skewness and kurtosis of the remaining unmasked coefficients. 

For our results to be directly comparable we perform the SMHW analysis for 
the same scales used by V03. For convenience and clarity,
the scales $R_i$ ($i$=1,2,...,15) plotted in the subsequent figures are listed in Table 1.
\begin{center}
Table 1\\
{\footnotesize{The scales used in the spherical mexican hat wavelet transform}}

{\footnotesize
\begin{tabular}{|llllllllllllllll|}
\hline 
 scales & $R_1$ & $R_2$ & $R_3$ & $R_4$ & $R_5$ & $R_6$ & $R_7$ & $R_8$ & $R_9$& $R_{10}$
 & $R_{11}$ & $R_{12}$ & $R_{13}$ & $R_{14}$ & $R_{15}$\\  
\hline 
 arcmin  & 14 & 25 & 50 & 75 & 100 & 150 & 200 & 250 & 300 & 400 & 500 & 600 & 750 & 900 & 1050 \\
\hline 
\end{tabular}
}
\end{center}

Fig. 1 shows our results, with the extended mask at each scale obtained by 
extending the $kp0$+sources mask such that all pixels closer than $2.5R$ to any 
of the pixels in the kp0+sources mask within $|b|<25^\circ$ are excluded from the 
analysis, attempting to follow closely the procedure of V03. The mask was 
not extended around point sources outside of this region as the $kp0$+sources mask around 
point sources seems to be in general extended enough to not cause contamination 
in wavelet coefficients on small scales, and on larger scales the effect gets 
averaged out.\footnote{There are however 3 bright semi-point sources in the northern 
hemisphere that are seen to visibly contaminate wavelet coefficients on scales 
less than $R_7$. These are taken care of by extending the mask out to $2.5R$ 
around these sources for scales less than $R_7$. There is also a more diffuse 
spot in the $kp0$ mask outside the $|b|<25^\circ$ region which causes visible 
contamination even on larger scales. The mask around this region is extended as well. 
The extended masks around these 4 regions are always retained even when the remaining 
details of the mask are varied. For reference, these regions are shown mapped in fig. 2 of 
Eriksen et al. 2004. The actual number of pixels in the coadded map that contain 
emission from these 4 sources together is quite small (less than 40, for $n_{side}$=256).}  
Conservatively extending the region around point sources too would leave too few pixels 
on scales of interest here. (It is unlikely that the non-Gaussianity signal found below, on 
$\sim4^\circ$ scales, is coming from point sources; see also Fig. 4(b) and related text in Sec.2.1).

The mean, $1\sigma$, $2\sigma$ and $3\sigma$ confidence contours obtained from 1000 
Gaussian simulations processed in the same way as the data are also shown in Fig. 1. The Gaussian simulations were created in the following way. CMB realizations with the same $C_l$ spectrum as the flat-$\Lambda$ CDM cosmology with power-law primordial power spectrum that best fits the 1st year WMAP data (Spergel et al. 2003; Hinshaw et al. 2003) were created at Healpix resolution $n_{side}=512$. Each realization was copied and smoothed with the WMAP beam window functions for each of the Q, V and W radiometer channels. Independent noize realizations of rms $\sigma_0/\sqrt{N_{obs}}$ were then added to the maps, where the effective number of observations $N_{obs}$ varies across the sky, and $\sigma_0$ is different for each radiometer channel. The 8 maps thus produced were coadded weighted by $N_{obs}/\sigma_0^2$. Thereafter the same analysis procedure that is applied to the data map is applied to each of the Gaussian simulations.  
 
If we estimate the significance of the signal using the generic $\chi^2$ test, that includes information on all scales,
\begin{equation}
\chi^2=\sum_{R_i,R_j} [S(R_i) - \bar{S}(R_i)] \Sigma^{-1}_{R_i,R_j} [S(R_j) - \bar{S}(R_j)],
\end{equation}
where $S(R_i)$ is the skewness or kurtosis signal on scale $R_i$, $\bar{S}(R_i)$ is the mean obtained from Gaussian simulations, and $\Sigma_{R_i,R_j}$ the covariance matrix obtained from simulations, we find that 
by comparing the $\chi^2$ of the signal in the data with the distribution of $\chi^2$'s obtained from Gaussian realizations, the kurtosis signal in 9 of 1000 realizations 
have larger $\chi^2$'s than the data. Hence we arrive at a significance of 99\% for this signal. The skewness signal is consistent with Gaussianity. Significances obtained using the $\chi^2$ test are tabulated in Table 2, under mask 1.

We find good consistency with the signal reported in V03. The 
kurtosis signal on $3-4^\circ$ scales in the southern Galactic 
hemisphere is outside the 
$3\sigma$ confidence contour. Only 2 and 3 simulations out of 1000 lead to a 
stronger kurtosis signal on scales $R_7$ and $R_8$ respectively, and only 1 of 
the Gaussian realizations has a larger kurtosis than the data on both the $R_7$ 
and $R_8$ scales in the southern hemisphere. Hence the signal on these particular scales appears significant at the 99.9\% confidence level. 

Even if the data were Gaussian there is a certain probability of 
obtaining outlier signals in its kurtosis spectrum on at least two of
the 15 scales considered. We take a closer look at 
this probability to better understand the significance of 
the detection. The number of Gaussian realizations that have kurtosis values that fall 
outside the 99\% confidence region in the southern hemisphere in $any$ two of the 15 scales is 28 (this 
number is 17 for a positive kurtosis), indicating that the signal found above on scales $R_7$ and $R_8$ in the data is significant 
at $at$ $least$ the 97\% confidence level. (The two scales were always consecutive in the simulations, but well 
spread out amongst all the scales.) Further these numbers are 22 and 17 for the northern hemisphere, and in none of these cases did the signal on both hemispheres lie outside their 99\% limits implying that a significant north-south asymmetry in the kurtosis signal in $any$ two scales was seen in 50 (37 for positive kurtosis) of 1000 Gaussian realizations. Thus the north-south asymmetry itself appears significant at $at$ $least$ the 95\% level.

Histogram plots of the wavelet coefficients on scale $R_7$ are shown in Fig. 2, 
for the all sky, northern and southern Galactic hemispheres. A longish tail towards negative 
values is seen in the southern hemisphere. Fig. 3 shows statistics relating to the 
minima, maxima (both in units of $\sigma$ on each scale) and $\sigma$ of the wavelet coefficients in the southern hemisphere, against scale, 
in the top panel. We use $\sigma$ to denote the $rms$ dispersion of the wavelet coefficients on each scale, noting
that 1$\sigma$, 2$\sigma$, etc., may not correspond 
to the same confidence levels as for a Gaussian distribution.
The middle and bottom panels show the statistics relating to the 
number of wavelet coefficients that were larger than 
(mean$+1\sigma$), (mean+2$\sigma$) 
and (mean+3$\sigma$) and smaller than (mean$-\sigma$), (mean-2$\sigma$) and (mean-3$\sigma$), 
respectively, again in the southern hemisphere. While the rest are seen to be quite 
consistent with limits obtained from Gaussian simulations, we see that the minima 
on $\sim 4^\circ$ scales is significant (with only $\sim 1\%$ of the simulations 
showing a stronger minima on each of scales $R_8$ and $R_9$), and the number of 
wavelet coefficients that are smaller than (mean-3$\sigma$) is very significant, 
with none of the simulations showing a stronger deviation on scales $R_6$ and $R_7$. 
This last estimator clearly gives a very strong signal. Only 3 of 1000 Gaussian realizations 
give a value for this estimator that is larger than the 99\% confidence contour in 
$any$ 4 scales. The signal we have here thus appears more significant than 99.7\% as 
in the data the value of this estimator on 4 scales is well out of the 99\% confidence region. The $\chi^2$ test described earlier also gives a similar significance for this signal. The number of cold pixels in the southern hemisphere is too large, on scales of $3-4^\circ$. On scales $R_6$ and $R_7$ there is more than one cold spot contributing to this number, while on larger scales it is mainly the one cold spot located near ($b=-57^\circ$, $l=209^\circ$) pointed out in V03. This spot is present on scales $R_6$ and $R_7$ as well. The corresponding statistics for the northern hemisphere are well consistent with Gaussianity.

Hence the non-Gaussianity shows up in the southern hemisphere in the form of a kurtosis signal and a larger than expected number of cold pixels.

\begin{center}
Table 2\\
{\footnotesize{Significance of deviation of the skewness and kurtosis signals from Gaussianity using the $\chi^2$ test}}

{\footnotesize
\begin{tabular}{|llllll|}
\hline 
 masks & mask 1 & mask 2 & mask 3 & mask 4 & ILC \\
\hline 
 skewness  & 44\% & 13\% & 57\% & 79\% & 24\% \\
 kurtosis  & 99\% & 99\% & 98\% & 99\% & 95\% \\
\hline 
\end{tabular}
}
\end{center}

\subsection{Other Masks}
We have checked that the above results are unaltered upon using a variety of 
different masks. We now show the kurtosis signal in the southern hemisphere 
for a few different masks. The mean, $1\sigma$, $2\sigma$ and $3\sigma$ confidence 
contours are obtained from Gaussian simulations processed each time in the same way as the data. 

Fig. 4(a) show the result of using an extended mask that is made in the same 
way as for Fig. 1 above but by extending the boundaries of the whole $kp0$ 
(without sources) mask by 2.5R, and then adding the mask around point sources back in on scales smaller than $R_7$. The shape of the mask is then retained on 
all scales.\footnote{We have also checked that applying the $kp0$ mask before or 
after performing the SMHW transform (but before applying the extended masks) does 
not affect the signal. Also, if we apply the SMHW tranform to the $kp0$ mask, and identify 
the wavelet coefficients that are say more than 1\% contaminated (by the edge of 
the mask), and include these pixels to make the extended masks, we have checked 
that the signal remains unaffected.} 

In Fig. 4(b) we show the result of using extended masks that on each scale apply a straight $|b|<(25+2.5R)^\circ$ 
Galactic cut, as well as a mask around point sources for scales smaller than $R_7$. The signal is thus unaffected by the shape of the mask. Fig. 4(c) 
shows the result of this time using a straight $|b|<(35+2.5R)^\circ$ Galactic cut. In 
going from Fig. 4(a) to 4(c) more of the sky is being excluded by the extended 
mask, and it is seen in the form of increased variance (this effect being larger 
for larger scales). But while the significance of the kurtosis signal seems to go 
down in this way, Fig. 5 shows that the number of cold pixels are in fact equally 
or more significant for this last mask.

Fig. 4(d) shows results from the ILC map. In this case, since there is little 
contamination from the Galactic plane left in the map, we can use just the $kp0$+sources 
mask without any extensions and apply it after the SMHW transform. The Gaussian 
simulations here were obtained by simulating the signal in each of the 10 radiometer 
channels, then smoothing to $1^\circ$ resolution, obtaining the noise weighted averaged 
signal for the 5 frequency channels, and then taking a linear combination of these with 
weights given in $\S$ 4 of Bennett et al. (2003). The signal is found in the ILC map too.

From the above analysis it is found that the kurtosis signal is indeed independent of the 
properties of the mask. The significances derived using the $\chi^2$ test for above masks are tabulated in Table 2. The masks corresponding to Figs. 4a,b,c are labelled mask 2, mask 3 and mask 4 respectively in the table; the ILC case with the unextended $kp0$+sources mask is labelled ILC.   

\subsection{Noise Simulations}
Finally Fig. 6 shows the kurtosis signal for the all sky case for the same mask as used in Fig. 1 but this time with confidence contours obtained from 110 Gaussian simulations that make use of the 110 full noise simulations provided by the WMAP team for each radiometer channel. The full noise simulations were made by generating one year of simulated time-series data including white noise, 1/f noise, and all inter-channel correlations that are known to exist in the radiometers, and then taking this data through all the steps of processing such as flight calibration, map-making, and filtering pipeline. We found that when compared with the case of 110 realizations of simple white noise for the same underlying sky simulations the two cases give identical results, so that when plotted simultaneously they are indistinguishable. This indicates that our simulations are reliable, and that the simple white noise model is completely satisfactory according to this statistic.

\section{Scale-scale Correlations}
 
Having obtained the wavelet coefficients of the data on different scales, we compute 
the scale-scale correlations between corresponding coefficients,
\begin{equation}
C_{R_i,R_j}= \frac{N \sum_{x} w(R_i,x)^2 w(R_j,x)^2}{ \sum_{x} w(R_i,x)^2 
\sum_{x} w(R_j,x)^2}.
\end{equation}
$w(R_i,x)$ are wavelet coefficients on scale $R_i$, and position or pixel $x$ in the sky. 
The coefficients that contribute to the sums are the $N$ unmasked coefficients on the 
larger scale. 

The top panel of Fig 7 shows scale-scale correlations between wavelet coefficients. The exact order of plotting is given in Table 3. Coefficients over the whole sky are used here. The mean, 1$\sigma$, 2$\sigma$ and 3$\sigma$ confidence contours obtained for the scale-scale correlations from Gaussian simulations are also shown. Consistency with Gaussianity is indicated, amongst these well separated scales.


Zooming into the scales that indicated non-Gaussianity in the previous section, the bottom panel of Fig. 7 shows the scale-scale correlations between coefficients of scales $R_6$ to $R_{11}$ in the southern Galactic hemisphere. The exact order of plotting is given in Table 3. Compared to Gaussian simulations we now find significant scale-scale correlations between the wavelet coefficients, presumably due to the cold spot that is common to all of these scales. A similar result is found for different masks.

\begin{center}
Table 3\\
{\footnotesize{The scales associated with the dummy indices plotted in Fig. 7}}

{\scriptsize
\begin{tabular}{|cccccccccccc|}
\hline 
index\# & 1 & 2 & 3 & 4 & 5 & 6 & 7 & 8 & 9 & 10
 & 11 \\  
top panel  & $R_1,R_2$ & $R_1,R_3$ & $R_1,R_5$ & $R_1,R_7$ & $R_1,R_{10}$ & $R_1,R_{14}$ & $R_2,R_3$ & $R_2,R_5$ & $R_2,R_7$ & $R_2,R_{10}$ & $R_2,R_{14}$ \\
\hline 
index\# & 12 & 13 & 14 & 15 & 16 & 17 & 18 & 19 & 20 & 21 & \\  
top panel  & $R_3,R_5$ & $R_3,R_7$ & $R_3,R_{10}$ & $R_3,R_{14}$ & $R_5,R_7$ & $R_5,R_{10}$ & $R_5,R_{14}$ & $R_7,R_{10}$ & $R_7,R_{14}$ & $R_{10},R_{14}$ & \\  
\hline
index\# & 1 & 2 & 3 & 4 & 5 & 6 & 7 & 8 & 9 & 10
 & 11 \\ 
bottom panel & $R_6,R_7$ & $R_6,R_8$ & $R_6,R_9$ & $R_6,R_{10}$ & $R_6,R_{11}$ &  $R_7,R_8$ & $R_7,R_9$ & $R_7,R_{10}$ & $R_7,R_{11}$ & $R_8,R_9$ & $R_8,R_{10}$ \\
\hline
index\# & 12 & 13 & 14 & 15 & & & & & & &  \\
bottom panel & $R_8,R_{11}$ & $R_9,R_{10}$ & $R_9,R_{11}$ & $R_{10},R_{11}$ & & & & & & &  \\
\hline
\end{tabular}
}
\end{center}

\section{$f_{NL}$ Constraints}
We showed in \S 2 that the skewness signal obtained using a range of masks was 
consistent with Gaussianity. In this section, we use the skewness signal on the 15 scales 
to compute the limits that can be placed on the amplitude of primordial non-Gaussianity, 
as parametrized by the non-linear coupling parameter, $f_{NL}$. Since the non-linear term (defined below) is not dominant and is quadratic, it is the skewness signal in the data, rather than any higher order cumulants, that is expected to be most sensitive to it. The kurtosis signal is not sensitive to this kind of non-Gaussianity (Cayon et al. 2003). 

$f_{NL}$ characterizes the amplitude of a quadratic term added to the curvature perturbations,
\begin{equation}
\Phi(x)=\Phi_L(x)+f_{NL}\left[ \Phi_L^2(x)-\langle \Phi_L^2(x)\rangle \right],
\end{equation}
where $\Phi_L$ are Gaussian linear perturbations with zero mean. Thus $f_{NL}$ 
parametrizes the leading order non-linear corrections to the primordial (curvature) 
perturbations.  The motivation to use data to place constraints on $f_{NL}$ is  
to address how Gaussian current CMB data are, or how much primordial non-Gaussianity, 
of this particular form, is allowed by the data. Such analyses also help compare 
the sensitivity of different data sets and of different estimators of non-Gaussianity to 
this particular form of non-Gaussianity.

Using an optimal estimator of non-Gaussianity based on the bispectrum, namely the cubic 
statistic, Komatsu et al. (2003) place limits of $-58<f_{NL}<134$ at 95\% confidence using 
the WMAP 1st-year data. They derive 95\% confidence limits of $f_{NL}<139$ based on Minkowski functionals. Smith et al. 2004, using VSA data, obtain an upper limit of 3100 at 95\% confidence (their limit being large because their data is sensitive to small scales).
Previously, using COBE DMR data, the bispectrum analysis of Komatsu et al. (2002) placed a 
limit of $|f_{NL}|<1500$, and using the skewness of SMHW coefficients Cay\'on et al. (2002) 
placed a limit of $|f_{NL}|<1100$, both at 68\% confidence. Using MAXIMA data Santos et al. 
(2002) placed a 1$\sigma$ limit of $|f_{NL}|<950$. 

We use the non-Gaussian simulations of Komatsu et al. (2003). How these have been produced 
are described in detail in the Appendix of their paper. Since producing these maps is a 
computationally intensive process, we use the 300 available realizations of non-Gaussian sky 
maps at HEALPix resolution $n_{side}$=256 to obtain the mean values of skewness at 
each of the different scales, for different values of $f_{NL}$. We use Gaussian simulations 
to estimate the covariance matrix of the skewness values for the different scales, and thus 
in turn to estimate the uncertainty in the measured $f_{NL}$. Since Gaussian simulations can 
be computed several orders of magnitude in time faster, we can estimate the covariance 
matrix accurately using a larger number of Gaussian simulations, and the uncertainty 
estimated from Gaussian simulations is a good approximation to that estimated from 
non-Gaussian simulations for $|f_{NL}|<500$\footnote{One way to see this is that the 1$\sigma$ uncertainty indicated by the 
different curves in figure 8 below are roughly the same indicating that at current sensitivity
of the data the uncertainties do not depend on the precise best fit
 value of $f_{NL}$.}.

In order to estimate the maximum likelihood $f_{NL}$, we compare the skewness values of 
the data at the 15 scales with simulations and use the goodness of fit statistic
\begin{equation}
\chi^2=\sum_{R_i,R_j} \left[ S(R_i)-\bar{S}_{sim}(R_i)\right] 
\Sigma_{R_i,R_j}^{-1} \left[ S(R_j)-\bar{S}_{sim}(R_j)\right],
\end{equation}
where $S(R_i)$ is the skewness of WMAP data on the ith scale $R_i$, 
$\bar{S}_{sim}(R_i)$ is the mean value from Monte Carlo simulations, 
computed for different values of $f_{NL}$, and $\Sigma_{R_i,R_j}$ 
is the scale-scale skewness covariance matrix from simulations. 

We have tested that the $\chi^2$ statistic accurately recovers $f_{NL}$ by using it on simulated maps. This is illustrated in Figure 8. The figure shows the mean $\chi^2$ distributions obtained from 300 simulated realizations of non-Gaussian maps with $f_{NL}$ values of 50, 100 and 150. The simulations include noise and window functions of the WMAp 1-yr data in the same as described earlier.

A plot of $\chi^2$ values obtained using data for different $f_{NL}$ is shown in figure 9.\footnote{This $\chi^2$ plot is obtained from using a diagonal covariance matrix of the skewness values on different scales. 1000 Gaussian simulations may be too small to obtain convergence for the off-diagonal elements of the covariance matrix. However, as also discussed in Eriksen et al. 2003b, if we are interested in obtaining the probability of the data given the Gaussian hypothesis, as was done in \S 2 of this paper, valid results can be obtained if we compute both the $\chi^2$ of the data and of the Gaussian realizations using a diagonal covariance matrix. Similarly a diagonal covariance matrix can be used here when we are interested in the relative change in $\chi^2$ with respect to a parameter. Although we do obtain consistent limits on $f_{NL}$ upon using the full covariance matrix.} $f_{NL}$ is thus estimated to be $50\pm80$ at 68\% confidence, with 95\% and 99\% upper limits of 220 and 280 respectively.

The limits on $f_{NL}$ can also be checked using the fisher discriminant function (Barreiro \& Hobson 2001, Cay\'on et al. 2003). 
An optimal linear function of the measured variables (here, skewness of wavelet coefficients on 15 different scales) is 
\begin{equation}
t({\bf{x}})=({\bm{\mu}}_0-{\bm{\mu}}_1)^T {\bf{W}}^{-1} {\bf{x}}.
\end{equation}  
Here {\bf{x}} is the 15 element vector that contains the skewness values at the different scales considered here, and $t({\bf{x}})$ is the fisher discriminant function that optimally puts together information contained in {\bf{x}} in the sense of maximizing the difference between the expected mean values of $t$ from Gaussian and non-Gaussian models, and minimizing their dispersions. ${\bf{W}}={\bf{V}}_0+{\bf{V}}_1$, the sum of the covariance matrices of the test statistic in the Gaussian (subscript 0) and non-Gaussian (subscript 1) cases. {\mbox{\boldmath{$\mu$}}}$_0$ and {\mbox{\boldmath{$\mu$}}}$_1$ are vectors containing the mean values of the test statistic in the Gaussian and non-Gaussian cases respectively.

Thus for different $f_{NL}$, the fisher discriminant function is found for each of the Gaussian realizations, for each of the non-Gaussian realizations, and for the data. The probability that the data are drawn from one or the other hypothesis can then be estimated. Looking at the fraction of non-Gaussian simulations that have values larger (for positive $f_{NL}$) and smaller (for negative $f_{NL}$) we deduce similar limits on $f_{NL}$ as derived above using the $\chi^2$ test. Barreiro \& Hobson (2001) found that the fisher discriminant can do better than $\chi^2$'s at distinguishing between Gaussian and non-Gaussian hypotheses. However since we have only 300 non-Gaussian simulations to obtain the fractions from, the accuracy of this method is not better but just comparable for our case here. Figure 10 shows the histograms of the fisher discriminants of 1000 Gaussian realizations, of 300 non-Gaussian realizations and of the data, for $f_{NL}$ values of 120 and 250; these values are close to the  the 1$\sigma$ and 2$\sigma$ limits derived using the $\chi^2$ above. For these values of $f_{NL}$, 0.74 and 0.95 of the non-Gaussian simulations, respectively, have larger values of the fisher discriminant than the data. We see that results obtained using fisher discriminants are consistent with those obtained using the $\chi^2$ test\footnote{For reference, the statistical power of the fisher discriminant test, as defined for example in Barreiro \& Hobson (2001) is found to be 0.28, 0.66, 0.88, and 0.98 for $f_{NL}$ values of 120, 250, 350, and 500,respectively, at the 95\% confidence level. These fractions are approximate as they have been obtained using only 300 non-Gaussian realizations.}.

We note that if the skewness signal in the data showed deviations from Gaussianity on particular scales then this method involving wavelet transforms could be used to obtain scale dependent constraints on $f_{NL}$. However the skewness signal in the data is well consistent with Gaussianity. The skewness spectrum of Fig 1(a) really flattens out with the use of another mask, such as the one used in Fig 4(a), while the kurtosis signal remains unchanged. Thus we do not believe there is reason to distinguish between different scales in obtaining constraints on $f_{NL}$ here. Also, the constraints on $f_{NL}$ derived here could possibly be made more stringent if we used the above method on the Wiener filtered map of primordial perturbations as discussed in Komatsu, Spergel \& Wandelt (2003). We will explore this in a future paper.

\clearpage

\begin{figure}
\plotone{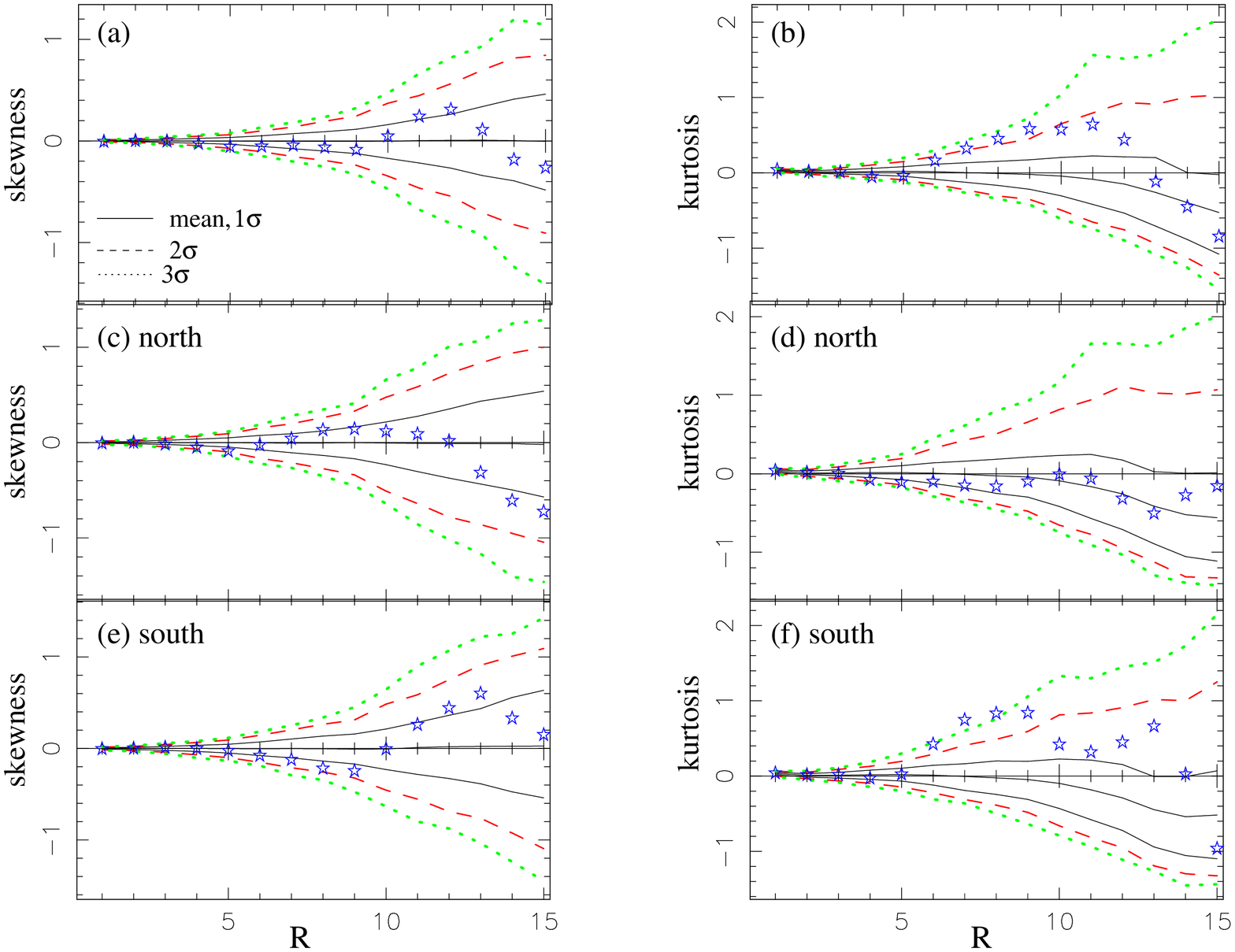}
\figcaption{Skewness and kurtosis spectra obtained using a SMHW analysis of the WMAP coadded data (stars), together with the mean values, and 1$\sigma$ (solid), 2$\sigma$ (dashed) and 3$\sigma$ (dotted) confidence contours obtained from 1000 Gaussian simulations. The top panel shows these for all sky, the middle for the northern Galactic hemisphere, and the bottom panel shows these for the southern Galactic hemisphere. Similar results were presented in V03.}
\end{figure}

\begin{figure}
\epsscale{0.8}\plotone{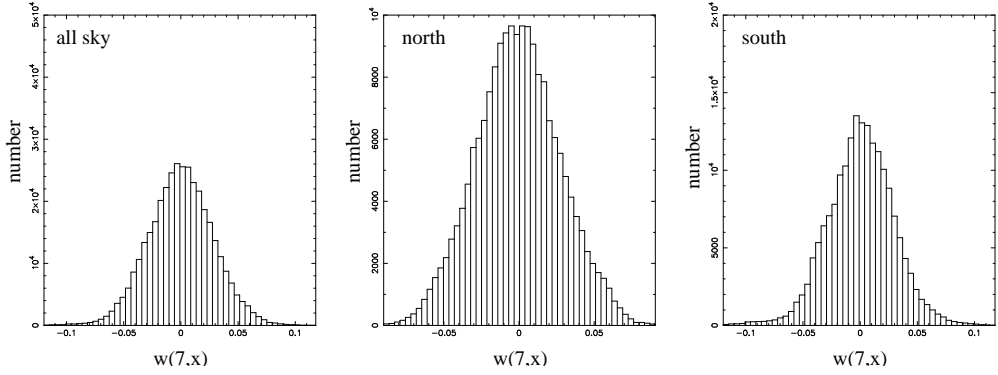}
\figcaption{Histograms of wavelet coefficients on scale $R_7$, for the all sky, northern and southern Galactic hemispheres.}
\end{figure}

\begin{figure}
\epsscale{1.05}\plotone{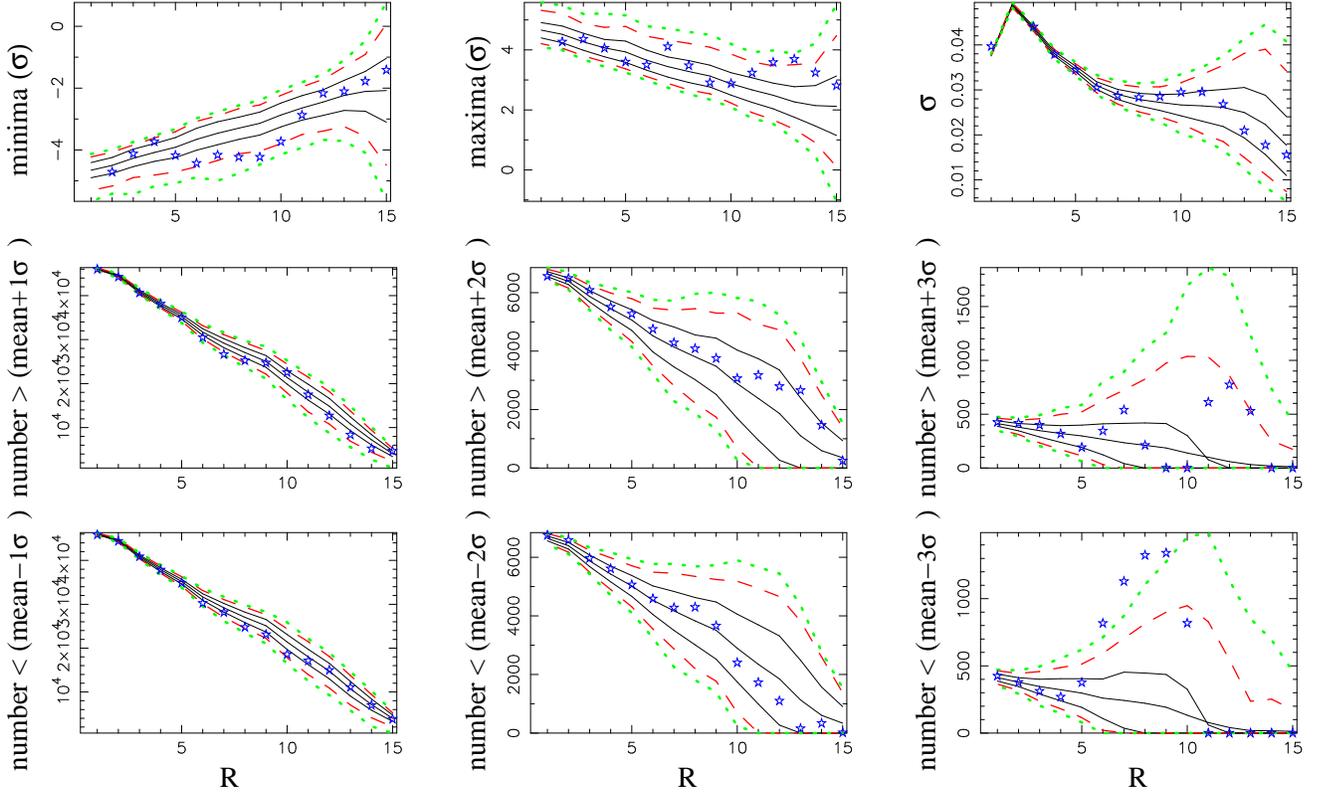}
\figcaption{Statistics relating to the minima and maxima, in units of $\sigma$, and $\sigma$ itself (top panel), and the number of wavelet coefficients that are larger than (mean$+\sigma$), (mean+2$\sigma$) and (mean+3$\sigma$) (middle panel), and smaller than (mean$-\sigma$), (mean-2$\sigma$) and (mean-3$\sigma$) (bottom panel), are shown, all for the southern Galactic hemisphere. Comparing the data (coadded WMAP; stars) with the confidence limits obtained from Gaussian simulations (mean and 1$\sigma$: solid, 2$\sigma$: dashed, and 3$\sigma$: dotted), the data show an excess of cold coefficients.}
\end{figure}

\begin{figure}
\epsscale{0.95}\plotone{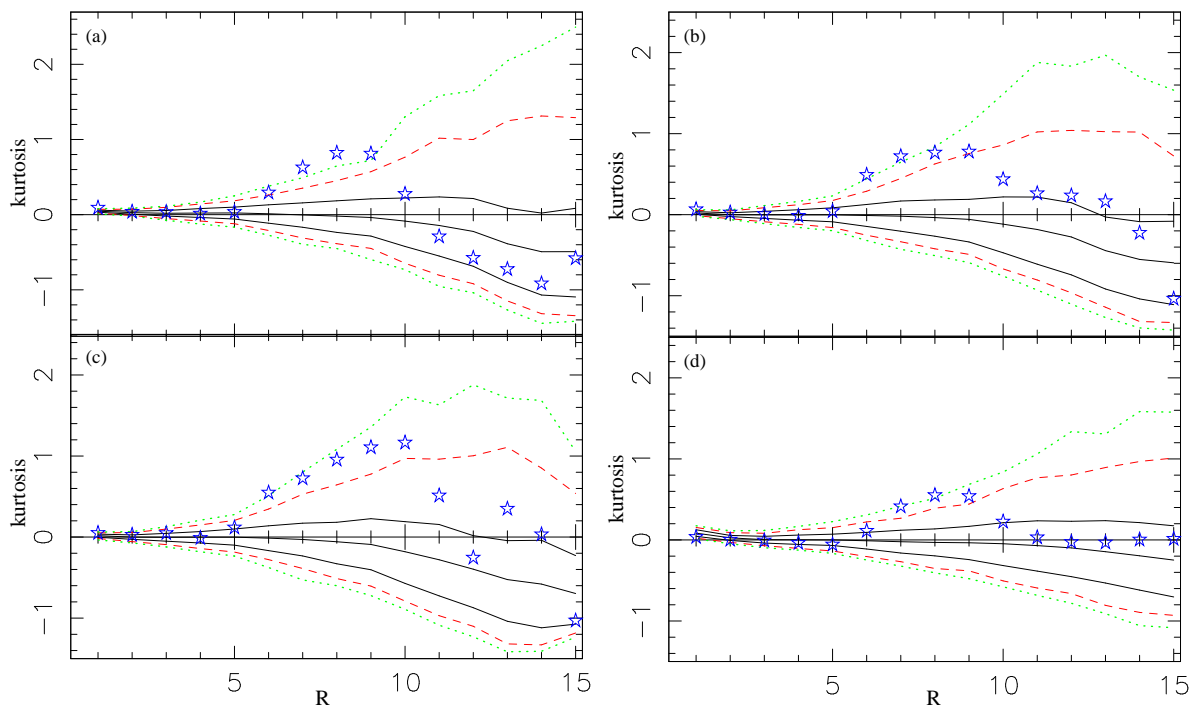}
\figcaption{Kurtosis spectra of wavelet coefficients in the southern Galactic hemisphere for different masks, as described in text.}
\end{figure}

\begin{figure}
\epsscale{1.0}\plotone{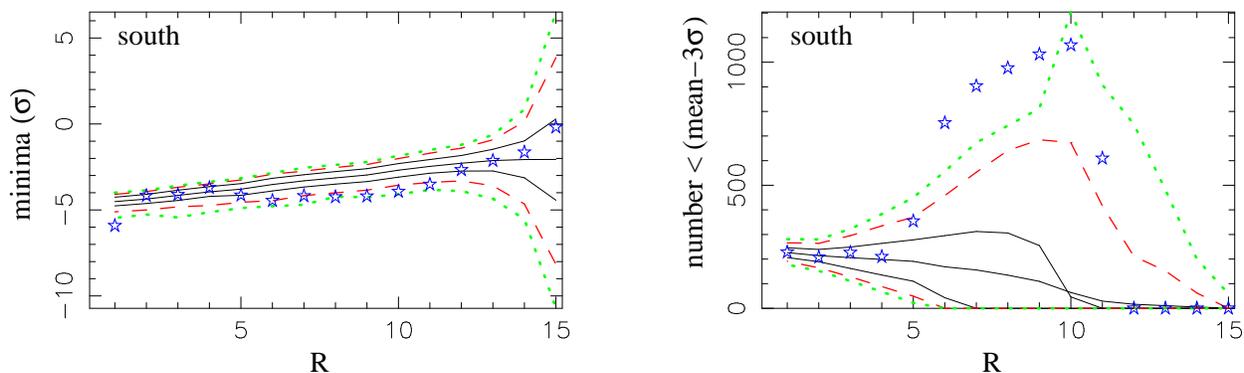}
\figcaption{Statistics relating to the minima, shown in units of $\sigma$, and to the number of wavelet coefficients that are smaller than (mean-3$\sigma$), in the WMAP coadded data (stars) in the southern Galactic hemisphere for a mask that is a straight $|b|<(35+2.5R)^\circ$ Galactic cut (see text) are shown. Also shown are the mean and $1\sigma$ (solid), 2$\sigma$ (dashed) and 3$\sigma$ (dotted) confidence contours obtained from Gaussian realizations.}
\end{figure}

\begin{figure}
\centering\includegraphics[angle=270,width=9cm]{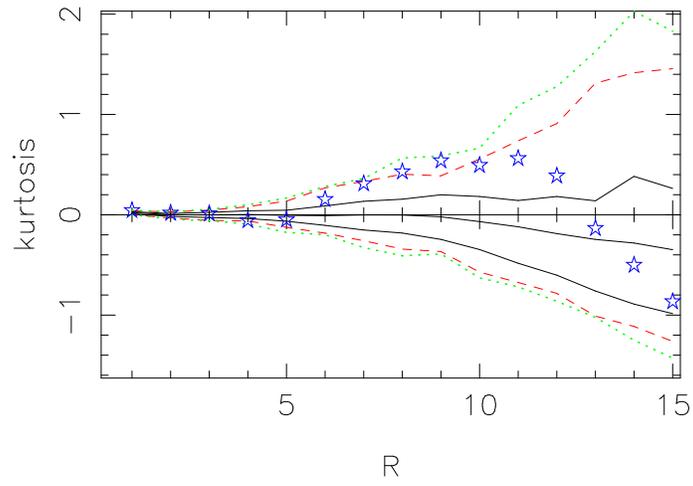}
\figcaption{This plot is to be compared with Figure 1(b). The confidence contours here have been obtained from 110 simulations using the 110 full noise simulations for each radiometer channel made available by the WMAP team.}
\end{figure}

\begin{figure}
\plotone{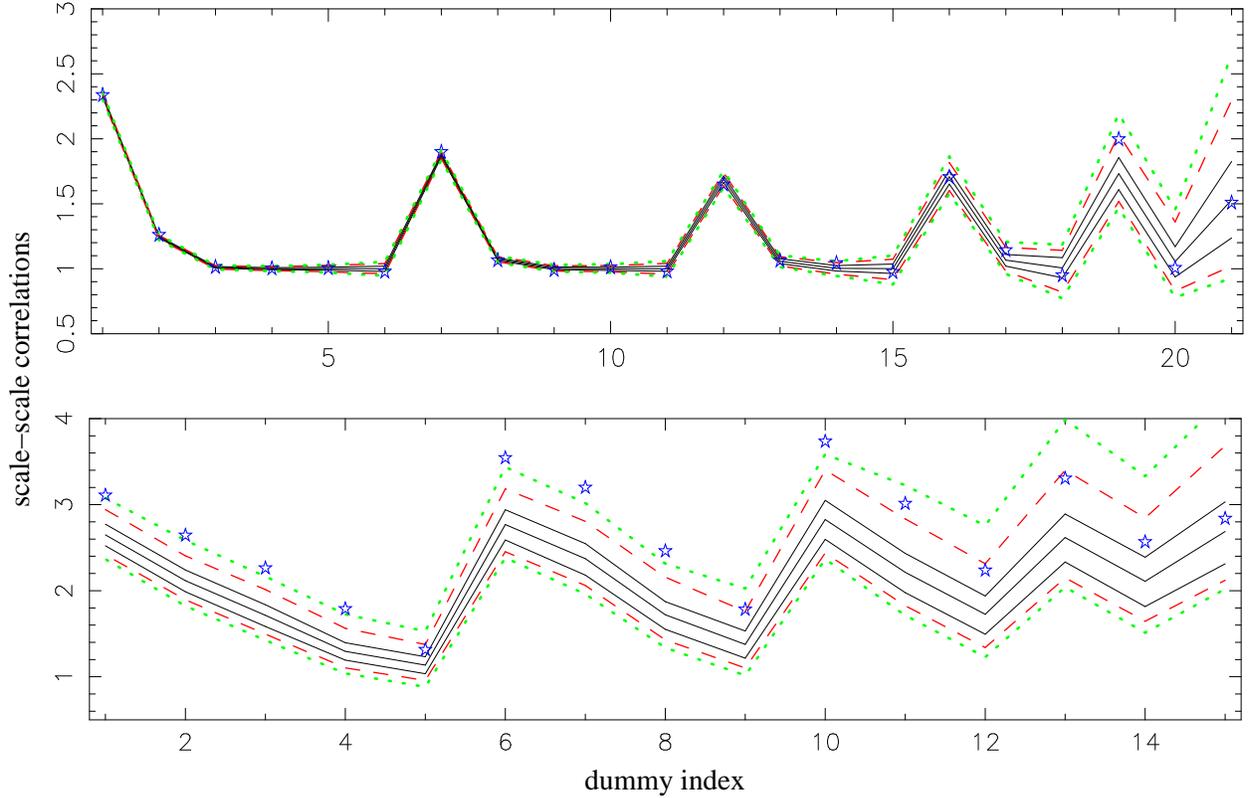}
\figcaption{The top panel shows scale-scale correlations amongst wavelet coefficients between scales that are well separated and span the whole range studied here. The bottom panel looks at scale-scale correlations amongst wavelet coefficients in the southern Galactic hemisphere between scales $R_6$ to $R_{11}$, the scales that show excess kurtosis and an excess of cold coefficients. See Table 3 for the exact order of plotting. The correlations obtained from the WMAP coadded data are shown as stars. Also shown are the mean and 1$\sigma$ (solid), 2$\sigma$ (dashed) and 3$\sigma$ (dotted) confidence contours obtained from Gaussian realizations.}
\end{figure}

\begin{figure}
\epsscale{0.65}\plotone{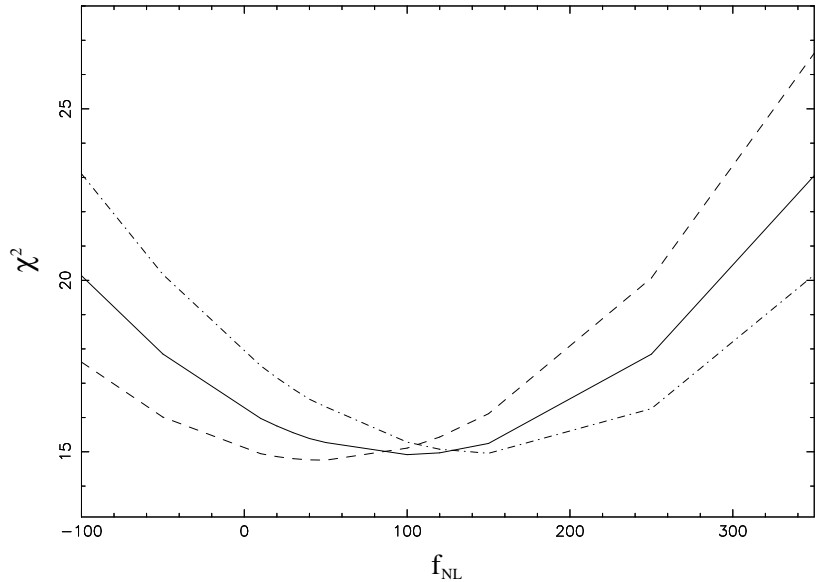}
\figcaption{Mean $\chi^2$ distributions obtained from 300 simulated realizations of non-Gaussian maps with $f_{NL}$ values of 50 (dashed), 100 (solid) and 150 (dot-dashed), illustrating that the statistic recovers the input value of $f_{NL}$. The simulations include as usual the noise properties and window functions of the WMAP 1-yr data. }
\end{figure}

\begin{figure}
\epsscale{0.65}\plotone{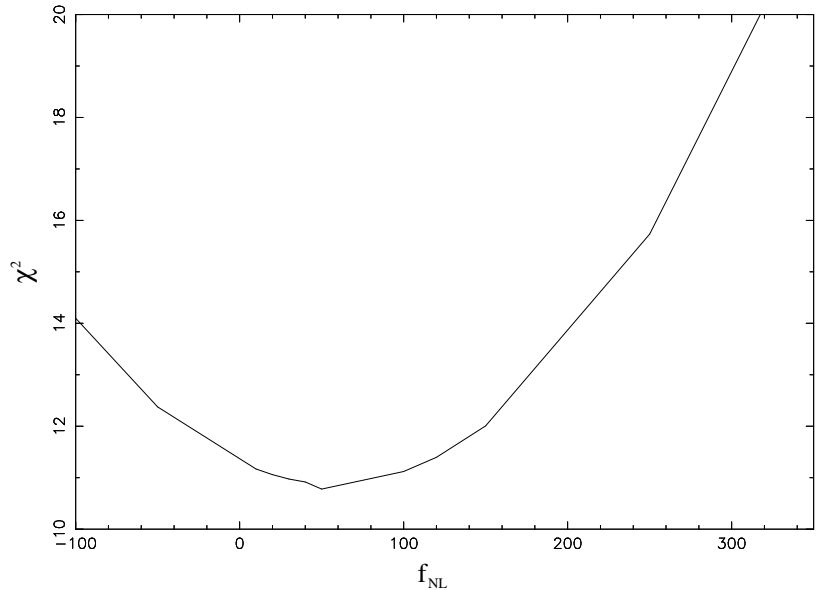}
\figcaption{A plot of $\chi^2$ against $f_{NL}$ obtained using WMAP data. $f_{NL}$ is thus estimated to be $50\pm80$ at 68\% confidence, and the 95\% and 99\% upper limits are 220 and 280 respectively.}
\end{figure}

\begin{figure}
\epsscale{0.95}\plotone{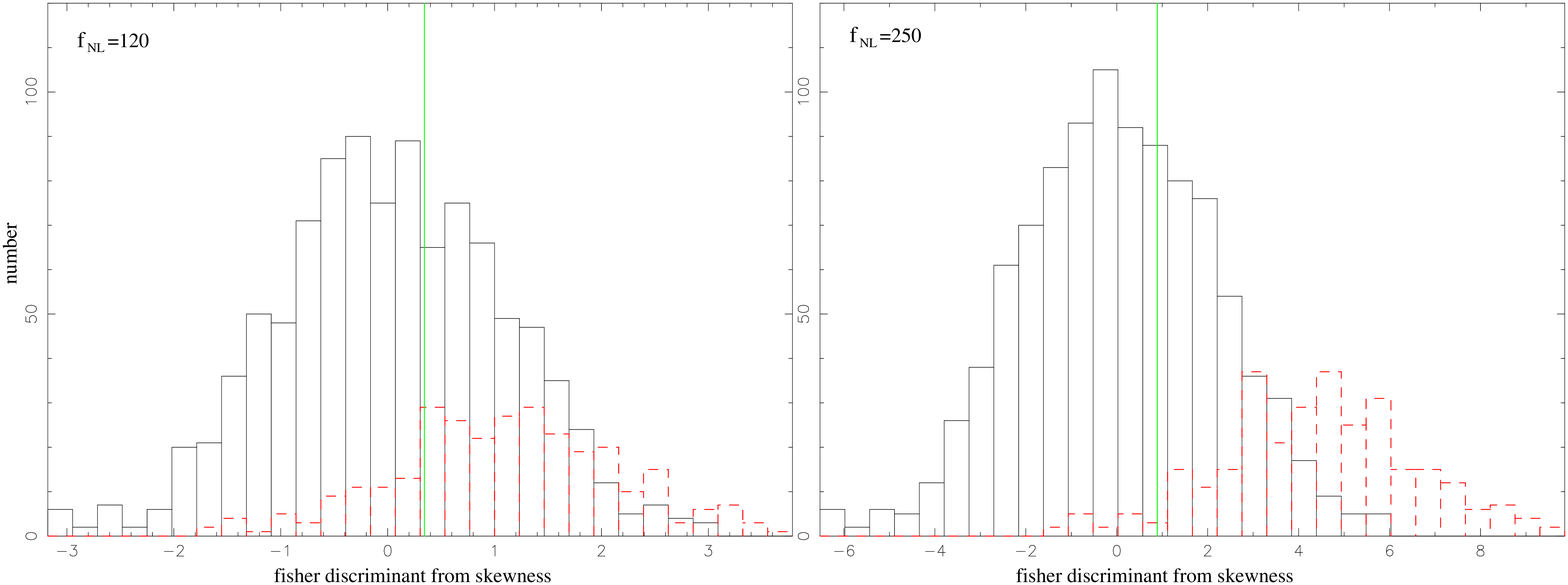}
\figcaption{Histograms of fisher discriminant values from 1000 Gaussian simulations (solid), from 300 non-Gaussian simulations (dashed), and the fisher discriminant of the data (vertical line), for $f_{NL}$ values of 120 and 250. These values are near the $1\sigma$ and 2$\sigma$ limits on $f_{NL}$. }
\end{figure}

\section{Conclusions}

We have analysed the first year WMAP data using a spherical mexican hat wavelet.
We detect non-Gaussianity at $\sim$ 99\% significance, consistent with that 
reported by Vielva et al. (2003). 
This detection corresponds to a positive kurtosis and to the presence of a larger than 
expected number of cold pixels (wavelet coefficients) in the southern Galactic 
hemisphere on scales $3-5^\circ$.

We have tested for changes in the significance of the signal with the type of mask 
used. The signal is found to be robust, and is found in the ILC map as well. We 
have also compared confidence contours obtained for the kurtosis spectra using 
the full noise simulation maps provided by the WMAP team, containing $1/f$ noise 
and other effects from data processing, to those obtained from using simulations 
that contain just white noise. We find very good agreement.

We have also applied another test statistic, the scale-scale correlations between 
wavelet coefficients. Significant scale-scale correlations are seen amongst the 
 coefficients over the range of scales that indicate the above non-Gaussianity.

We then use the skewness statistic on the different scales to place constraints 
on the non-linear coupling parameter $f_{NL}$, the motivation being to see how 
much non-Gaussianity of this particular form is allowed by current data. It 
is also a way to compare the sensitivity of different test statistics to this parameter. 
Constraints obtained are closely consistent with those obtained by Komatsu et al. (2003) using the cubic statistic and Minkowski functionals on the same data. The constraints on $f_{NL}$ derived here could possibly be made more stringent if we used spherical wavelets on the Wiener filtered map of primordial perturbations as discussed in Komatsu, Spergel \& Wandelt (2003). We will explore this in a future paper.
The kurtosis statistic is not sensitive to this form of non-Gaussianity.  
We will present constraints on other forms of non-Gaussianity implied
by the kurtosis statistic of the WMAP data elsewhere. 

\section{Acknowledgements}
We acknowledge the use of non-Gaussian simulations from Komatsu et al. 
(2003) [WMAP paper]. It is a pleasure to thank Eiichiro Komatsu for 
beneficial discussions and comments on the manuscript.  
PM thanks Gang Chen for useful discussions. We acknowledge the use of OSCER supercomputing facilities at the University of Oklahoma.
This work is supported in part by NSF CAREER grant AST-0094335. 

\section{Appendix}
The SMHW transform has been used previously by Cay\'on et al. (2001,2003); Mart\'inez Gonz\'alez et al. (2002), and V03. It is a continuous and symmetric wavelet which in the small angle limit corresponds to the euclidean mexican hat wavelet. The wavelet is given by
\begin{equation}
\Psi_R(y)=\frac{1}{\sqrt{2 \pi} N(R)} \left[1+(\frac{y}{2})^2\right]^2 \left[2-(\frac{y}{R})^2\right] e^{-y^2/2R^2},
\end{equation}
where $R$ is the scale and $N(R)$ is a normalization constant given by
\begin{equation}
N(R)=R (1+\frac{R^2}{2}+ \frac{R^4}{4} )^{1/2}.
\end{equation}
$y=2$tan$\frac{\theta}{2}$, for polar angle $\theta$. $\psi$ is thus isotropic so that when centered on the north pole it is independent of the polar coordinate $\phi$. Thus moving any point to the north pole, the wavelet coefficients 
\begin{equation}
w(R,x':(\theta',\phi'))=\int dx \, T(x) \, \Psi_R(|x'-x|)
\end{equation}
can be obtained via a convolution of the sky map $T(x:\theta,\phi)$ with the wavelet function. This is easily performed in spherical harmonic space.


\begin{thebibliography}{}


\bibitem[Aghanim et al.(2003)]{aghanim03}
Aghanim, N., Kunz, M., Castro, P.G., \& Forni, O. 2003, Astron.Astrophys., 406, 797

\bibitem[Bennett, et al.(2003)]{bennett03}
Bennett, et al., 2003, astro-ph/0302208 

\bibitem[Cabella et al.(2003)]{cabella03}
Cabella, P., Hansen, F., Marinucci, D., Pagano, D., \& Vittorio, N. 2003, astro-ph/0401307

\bibitem[Cayon et al.(2003)]{cayon03}
Cay\'on, L.; Mart\'inez-Gonz\'alez, E.; Arg\"ueso, F.; Banday, A. J.; G\'orski, K. M. 2003, MNRAS, 339, 1189

\bibitem[Cayon et al.(2001)]{cayon01}
Cay\'on, L.; Sanz, J. L.; Mart\'inez-Gonz\'alez, E.; Banday, A. J.; Arg\"ueso, F.; Gallegos, J. E.; G\'orski, K. M.; Hinshaw, G. 2001, MNRAS, 326, 1243

\bibitem[Chiang et al.(2003)]{chiang03}
Chiang, L.-Y., Naselsky, P.D., Verkhodanov, O.V., \& Way, M.J. 2003, ApJ, 590, 65

\bibitem[Colley, \& Gott(2003)]{colley}
Colley, W.N., \& Gott, J.R., 2003, MNRAS, 344, 686

\bibitem[Copi, Huterer, \& Starkman(2003)]{copi03}
Copi, C.J., Huterer, D., \& Starkman, G.D. 2003, astro-ph/0310511

\bibitem[Eriksen et al.(2003)]{eriksen03b}
Eriksen, H.K., Novikov, D.I., Lilje, P.B., Banday, A.J., Gorski, K.M. 2003, astro-ph/0401276

\bibitem[Eriksen et al.(2003)]{eriksen03a}
Eriksen, H.K., Hansen, F.K., Banday, A.J., Gorski, K.M., \& Lilje, P.B. 2003, astro-ph/0307507

\bibitem[Gaztanaga \& Wagg(2003)]{gaztanagawagg}
Gaztanaga, E., \& Wagg, J. 2003, Phys.Rev. D68, 021302

\bibitem[Gurzadyan et a.(2004)]{gurzadyan04}
Gurzadyan, V.G., et al. 2004, astro-ph/0402399

\bibitem[Hansen et al.(2004)]{hansen04}
Hansen, F.K., Cabella, P., Marinucci, D., \& Vittorio, N. 2004, astro-ph/0402396

\bibitem[Hobson, Jones, \& Lasenby(1999)]{hobson99}
Hobson, M. P., Jones, A. W., \& Lasenby, A. N. 1999, MNRAS, 309, 125

\bibitem[Barreiro, \& Hobson(2001)]{barreirohobson}
Barreiro, R. B., \& Hobson, M. P. 2001, MNRAS, 327, 813

\bibitem[Barreiro et al.(2000)]{barreiro00}
Barreiro, R. B., Hobson, M. P., Lasenby, A. N., Banday, A. J., G\'orski, K. M., \& Hinshaw, G.  2000, MNRAS, 318, 475

\bibitem[Komatsu, Spergel \& Wandelt(2003)]{komatsu03b}
Komatsu, E., Spergel, D.N., Wandelt, B.D. 2003, astro-ph/0305189 

\bibitem[Komatsu et al.(2003)]{komatsu03a}
Komatsu, E., et al. 2003, ApJS, 148, 119

\bibitem[Martinez-Gonzalez et al.(2002)]{martinez02}
Mart\'inez-Gonz\'alez, E.; Gallegos, J. E.; Arg\"ueso, F.; Cay\'on, L.; Sanz, J. L 2002, MNRAS, 336, 22

\bibitem[Mukherjee, Hobson, \& Lasenby(2000)]{mukherjee00}
Mukherjee, P., Hobson, M.P., \& Lasenby, A.N.  2000, MNRAS, 318, 1157

\bibitem[Pando, Valls-Gabaud, \& Fang(1999)]{pando}
Pando, J., Valls-Gabaud, D., \& Fang, L.-Z. 1999, Phys.Rev.Lett. 81, 4568

\bibitem[Vielva et al.(2003)]{vielva03}
Vielva, P., Mart\'inez-Gonz\'alez, E., Barreiro, R.B., Sanz, J.L., \& Cay\'on, L. 2003, ApJ in press, astro-ph/0310273 (V03)

\bibitem[Santos, et al.(2003)]{santos03}
Santos, M.G.,  et al. 2003, MNRAS, 341, 623 

\bibitem[Smith, et al.(2004)]{smith04}
Smith, S., et al. 2004,astro-ph/0401618

\bibitem[Starck, Aghanim, \& Forni(2003)]{starck03}
Starck, J.L., Aghanim, N., \& Forni, O. 2003, astro-ph/0311577



\end{thebibliography}
\end{document}